\documentclass[fleqn,10pt]{wlscirep}
\UseRawInputEncoding
\usepackage[utf8]{inputenc}
\usepackage[T1]{fontenc}

\title{Simultaneous optical trapping and magnetic micromanipulation of ferromagnetic iron-doped upconversion microparticles in six degrees of freedom}

\author[1]{Gokul Nalupurackal}
\author[1]{Gunaseelan M.}
\author[1]{Muruga Lokesh}
\author[1]{Rahul Vaippully}
\author[2]{Amit Chauhan}
\author[2]{B. R. K. Nanda}
\author[2]{Chandran Sudakar}
\author[3]{Hema Chandra Kotamarthi}
\author[4]{Anita Jannasch}
\author[4]{Erik Schäffer}
\author[5]{Jayaraman Senthilselvan}
\author[1,*]{Basudev Roy}
\affil[1]{Department of Physics, Quantum Centres in Diamond and Emergent Materials (QuCenDiEM)-group, Micro Nano and Bio-Fluidics (MNBF)-Group,
Indian Institute of Technology Madras, Chennai, India 600036}
\affil[2]{Condensed Matter Theory and Computational Lab, Department of Physics,
Indian Institute of Technology Madras, Chennai - 600036, India}
\affil[2]{Center for Atomistic Modelling and Materials Design, Indian Institute of Technology Madras, Chennai - 600036, India}
\affil[3]{Department of Chemistry, Indian Institute of Technology Madras, Chennai, India, 600036}
\affil[4]{Center for Plant Molecular Biology (ZMBP), University of Tübingen, Germany}
\affil[5]{Department of Nuclear Physics, University of Madras, Chennai, India, 600036}
\affil[*]{basudev@iitm.ac.in}

\begin{abstract}
Optical trapping of magnetic Fe-oxide particles is notoriously difficult due to their high refractive indices, not to mention high absorptivity at the trapping infra-red wavelengths. We synthesize Fe co-doped NaYF$_4$: Yb, Er ferromagnetic upconversion particles that not only have refractive indices conducive for optical trapping, but also heat less than Haematite particles at off-resonant wavelengths. These particles are hexagonal shaped with dimensions of the order of 3\,$\mu$m and also bear high coercivity of 20 mT and saturation magnetisation of 1 Am$^2$/kg. This enables simultaneous use of optical trapping and magnetic forces to generate micro-manipulation in all the six degrees of freedom of a rigid body. We also show that these particles heat significantly when illuminated on absorption resonance at 975 nm while emitting visible light with possible implications for fluorescence microscopy and photothermal therapy of cancer cells.  
\end{abstract}
\begin{document}

\flushbottom
\maketitle
%
%
\thispagestyle{empty}

\section*{Introduction}

The observation of physical and biological processes at the molecular scale requires high spatial and temporal resolution during the application of large forces and torques. In the biological applications, a variety of high-resolution single-molecule techniques, including single molecule fluorescence resonance energy
transfer (FRET) \cite{joo2008advances,holden2010defining}, optical tweezers \cite{cheng2011single,abbondanzieri2005direct,greenleaf2005passive}, magnetic tweezers  \cite{koster2005friction,strick2000single}  and total internal reflection fluorescence (TIRF)\cite{tirf1,tirf2} have been developed to observe the 
individual enzymatic steps \cite{dulin2013studying}. Optical tweezers implemented in the dumb-bell configuration has achieved the best spatiotemporal resolution yet in enzymatic studies\cite{kimura2006single}. The positions of two trapped microspheres linked together by a nucleic acid are controlled, allowing enzymatic activity on
the nucleic acid to be read out from the displacement of the particles from their equilibrium positions \cite{neuman2004optical}. But yet, the forces applicable with optical tweezers are typically limited to about 100\,pN unless special types of core-shell particles are used\cite{jannasch2012nanonewton}. Moreover, these are restricted to perform only in-plane modes of rotation. 

 Magnetic fields cannot create a stable stationary equilibrium configuration to confine a particle (Earnshaw’s Theorem). Special geometries have to be used to circumvent this, using a combination of magnetic and other forces \cite{ref10,ref11}, or sophisticated electromagnets \cite{ref12}. In these experiments, the molecule of interest is typically tethered between a surface and a superparamagnetic particle, and the force is applied via an external magnetic field. In this way, forces larger than 200\,pN and controlled torques in all senses can be applied \cite{bausch1999measurement,neuman2008single}. However, the inability to decouple force and torque application is a major impediment in magnetic tweezers.

Typical magnetic tweezers track the position of particles at an acquisition frequency of ~50–100\,Hz and resolve the subsequent position of a particle along the z axis to about 1\,nm using video cameras \cite{van2012non,cnossen2014optimized,gosse2002magnetic}. Recent improvements to magnetic tweezers based on high-speed complementary metal-oxide semiconductor (CMOS) cameras have enhanced the resolution to the sub-Angstrom level at 1\,s bandwidth on individual particles \cite{lansdorp2013high,huhle2015camera}. However, it can still be improved with higher bandwidth measurements, both in terms of speed and in accuracy, by using interferometric detection. 

Few of the early attempts at a combination magneto-optical tweezers used monodispersed polystyrene substrate coated in
a layer of chromium oxide (CrO$_2$), which was held together by an outer polystyrene layer \cite{photonics2030758} and Micromer magnetic beads (3 $\mu$m diameter, Micromod, Rostock, Germany) selected for their low magnetic content, permitting good optical trapping \cite{crut2007fast}. The chromium oxide shell material, although claimed to be ferromagnetic, loses this behavior at room temperatures, not to mention low magnetisation values. In all these particles, the presence of magnetic materials makes optical trapping complicated due to the high refractive index and high optical absorption. 

Here, we present, an alternative strategy where the tracer particle is not only highly magnetic but also easily optically trappable. This provides the best of magnetic forces in conjugation with high spatiotemporal detection that the optical tweezers provides via interferometric detection with the quadrant photodiodes.  We show that Fe-doped upconverting NaYF$_4$:Yb,Er particles (Fe-NYF) have ferromagnetic features and yet the refractive index is well conducive for optical tweezers measurements (the lattice NaYF$_4$ has a refractive index of 1.55) , even for particles as large as 5\,$\mu$m, with large saturation magnetisation of 1 Am$^2$/kg . The upconverting NaYF$_4$:Yb,Er (NYF) non-magnetic particles have been routinely used with optical tweezers \cite{sumitfrontinphys,kumar2020pitch,sumitnanoscale} but, for the first time, the magnetic properties of these NYF through co-doping with Fe are being explored. We show that this co-doped particle can be held in optical tweezers while the magnetic field could be used to turn the particle in all the three degrees of freedom, namely, yaw \cite{friese1998optical,rahulyaw,rahulsyaw2,basudevpnas}, pitch \cite{lokeshjopc,lokeshrscadv,rahulsoftmatter,kumar2020pitch} and roll, in the nomenclature of the airlines. 

We also show that the Fe-NYF's heat up upon illumination with infra-red light, particularly if the absorption resonance wavelength of 975 nm is used. It is generally believed that there is an apparent self-heating when these particles are used for thermometry \cite{pickel2018apparent}. We show that the effect of self-heating is actually real when the local region of the particle heats up significantly to either form water vapor bubbles or even kill cancer cells. It is also known that water itself is very absorptive at 975 nm, but such a beam does not form vapor bubbles when illuminated on the water without the particle. We also envisage that such materials can be used as microrobots to manoeuvre the particle to the desired location under the influence of magnetic fields where a laser beam at the absorption resonance wavelength would heat the particle. This can be helpful for photothermal therapy.

\section*{Results}

\subsection*{The Material}

\begin{figure}[h]
	\centering
		\includegraphics[width=0.7\linewidth]{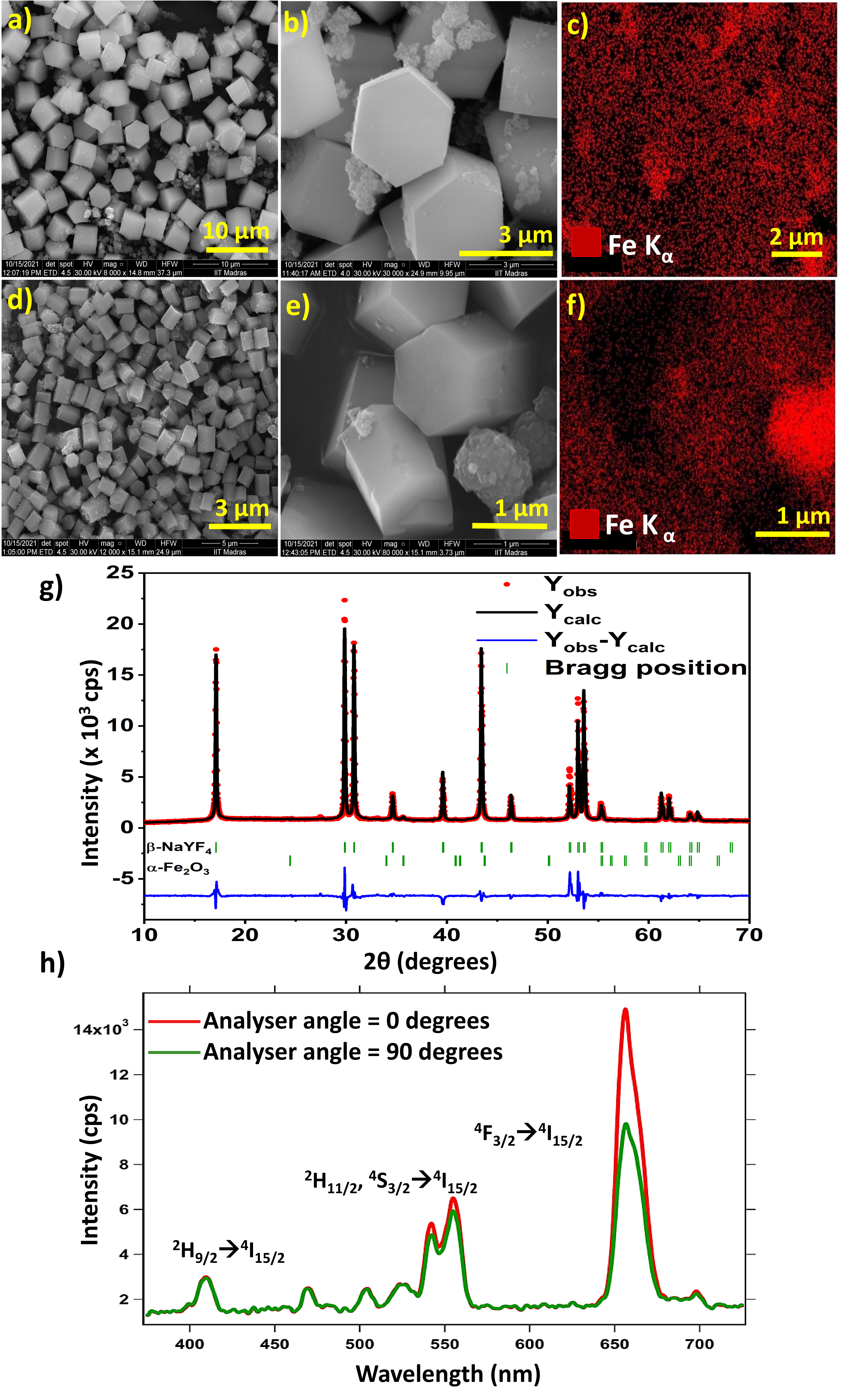}
	\caption{The FE-SEM images of Fe-15 and Fe-30 NYF upconverting particles are shown in figures(a-b) and (d-e) respectively. Figures(c) and (f) represents the elemental mapping of iron corresponding to figures (b) and (e) respectively. The simulated and observed XRD patterns of Fe-15 NYF(g), and the emission spectrum(h) recorded at orthogonal angles of a polariser with respect to the side-on axis of the particle are also shown when illuminated by 975 nm light. When a hexagonal UCP is optically trapped, it prefers to orient with the side-on axis facing the tweezers beam.}
	\label{sem}
\end{figure}

The Fe doped NaYF$_4$:Yb,Er upconverting particles (Fe-NYF) with sizes larger than 2 $\mu$m were prepared using hydrothermal process. The field emission scanning electron microscope (FE-SEM) images and the corresponding energy dispersive x-ray spectroscopy (EDS) elemental maps for the particles with different degrees of doping (15 at.\% and 30 at.\%, Fe-15NYF and Fe-30NYF, respectively) are shown in Fig. \ref{sem}(a-f). The Fe-NYF are formed as precise hexagonal shaped particles with different sizes. The estimated average side length $\times$ width are 3.2 $\mu$m $\times$ 2.8 $\mu$m and 1.3 $\mu$m $\times$ 1.4 $\mu$m  for Fe-15 and Fe-30 NYFs, respectively. The EDS element mappings were performed to confirm the presence of Fe in the crystal lattice. The Fe K$_{\alpha}$ elemental maps from the regions shown in Fig.\ref{sem}(b and e) are shown in Fig.\ref{sem}(c and f). These elemental maps reveal the existence of uniformly distributed Fe in the hexagonal Fe-NYF’s. In addition, few Fe-rich particles with no definite morphological feature, mostly from $\alpha$-Fe$_2$O$_3$ impurities, are seen lying next to these Fe-NYF’s.

Fig.\ref{sem}(g) shows the X-ray diffraction (XRD) pattern (red dots) for the Fe-15 NYF. The Rietveld refinement performed using the FullProof software with the simulated pattern and obtained with a best $\chi^2$ is shown in black line \cite{rodriguez2001fullprof,young1993rietveld}. The prepared samples crystallize in hexagonal structure of NaYF$_4$  (\(\beta\)-NaYF$_4$) with the space group of P6$_3$/m. Despite the substitution of Fe$^{3+}$ at the Y$^{3+}$ site the structure remains in the hexagonal phase. The estimated lattice parameters a=5.9813(4) and b=3.5052(7) matches well with the structure of $\beta$-NaYF$_4$ (JCPDS No.: 16-0334). In addition to the $\beta$-NaYF$_4$, the samples exhibit small phase fraction of $\alpha$-Fe$_2$O$_3$ impurity (0.96 wt.\%). If Fe does not go into the lattice of $\beta$-NaYF$_4$ structure, the diffraction pattern would show ~13 wt.\% $\alpha$-Fe$_2$O$_3$ impurity phase. Fe-30NYF samples also show $\beta$-NaYF$_4$ structure, however, with higher percentage of $\alpha$-Fe$_2$O$_3$ (2.5 wt.\%)​phase. For the present study we have used Fe-15NYF's as it exhibits optimal magnetization values need for the optical trapping experiments.

\begin{figure}[h]
	\centering
		\includegraphics[width=0.6\linewidth]{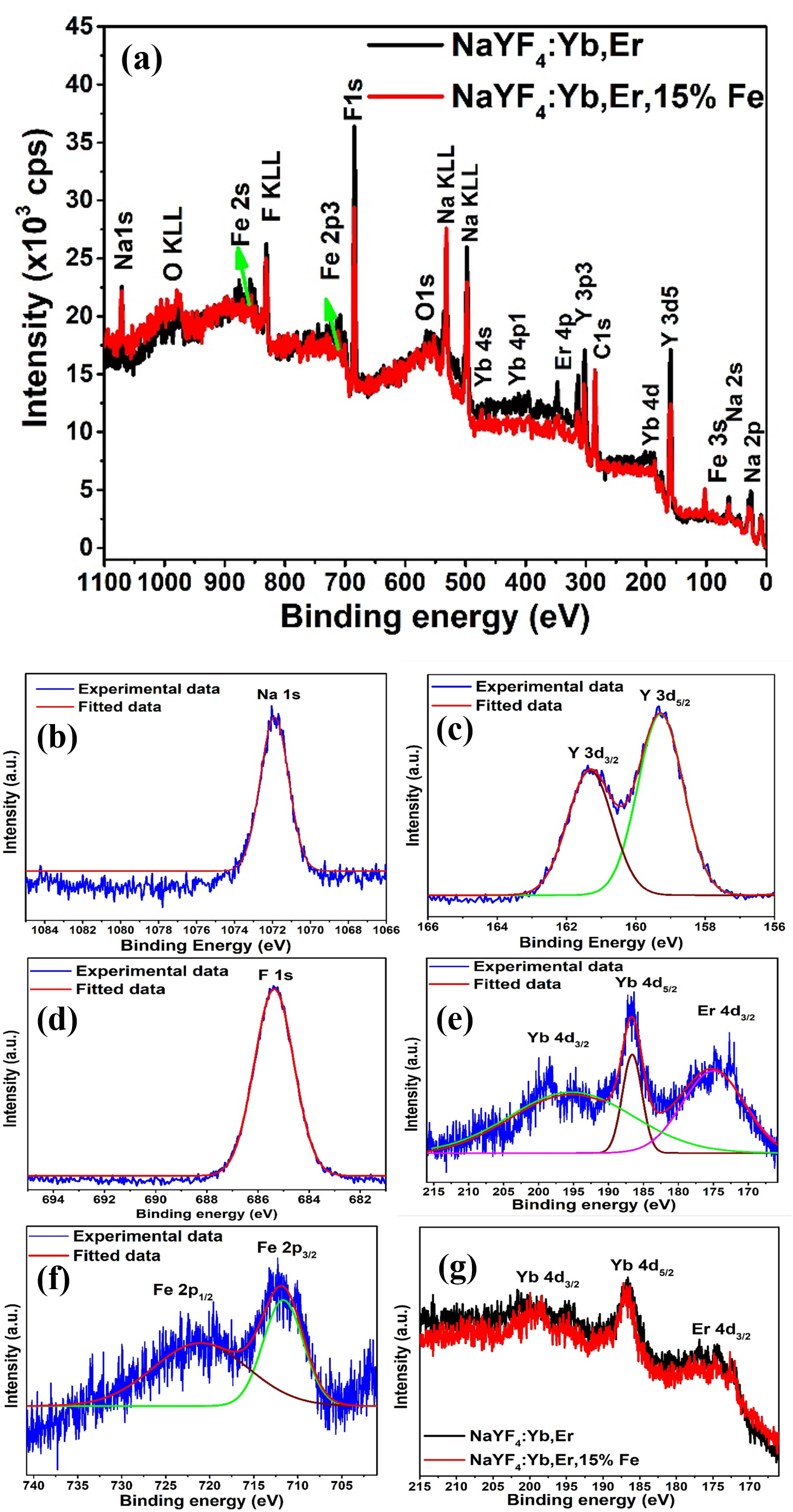}
    \caption{XPS spectra of Fe-15 NYF is shown in the above figure. In (a) Survey spectra, (b) Na 1s, (c) Y 3d, (d) F 1s, (f) Fe 2p and (e,g) Yb 4d, Er 4d. are shown.}
    \label{xps}
\end{figure}

The X-ray photoelectron spectroscopy (XPS) was used to confirm the valency of elements in the Fe-NYF's and estimate the at.\% of elements in the compound. A representative analyses of XPS spectra performed on NaYF$_4$:Yb,Er NYF's and Fe-15 NYF is shown in the Fig.\ref{xps}. The survey spectrum shown in Fig.\ref{xps}(a) clearly shows the presence of Na, Y, F, Yb and Er elements in the sample. In addition, XPS signals from dopant Fe are clearly discerned. The strong signals from Na 1s, Y 3d and F 1s spectra shown in Fig.\ref{xps}(b-d) confirm the formation of NaYF$_4$. Further, the presence of Yb and Er 4d spectra lines are detectable despite the signal arising from few tens of Angstrom layers of the surface region as shown in Fig.\ref{xps}(e). The binding energy of photoelectrons from all these elements, viz Na (+1), Y (+3), F (-1), Yb (+3) and Er (+3), suggests that the oxidation states are consistent with the known composition. Further the integral intensity used to estimate the composition also confirm the nominal composition taken in the synthesis. The 2p photoelectron spectra from the element Fe is indicating the +3 oxidation state with Fe 2p$_3$$_/$$_2$ peaks appearing at ~711 eV. Due to the oxidation state and favorable ionic size we expect that the Fe atoms mostly occupy Y sites which is also occupied by the Yb and Er dopants. The estimates on the Fe concentration is also consistent with the nominal concentration taken during the synthesis. The high resolution spectral features of the elements in the pristine NaYF$_4$:Yb,Er(NYF) and Fe doped NaYF$_4$:Yb,Er (Fe-NYF) are similar, specifically shown in Fig.\ref{xps}(g) for Yb and Er XPS spectra. This strongly suggests that Fe-15 NYF particles are compositionally similar to the pristine compounds of NaYF$_4$:Yb,Er and that preferably Fe ions occupy Y$^{3+}$ sites as it has 3+ valence state.

\begin{figure}[ht!]
	\centering
		\includegraphics[width=0.5\linewidth]{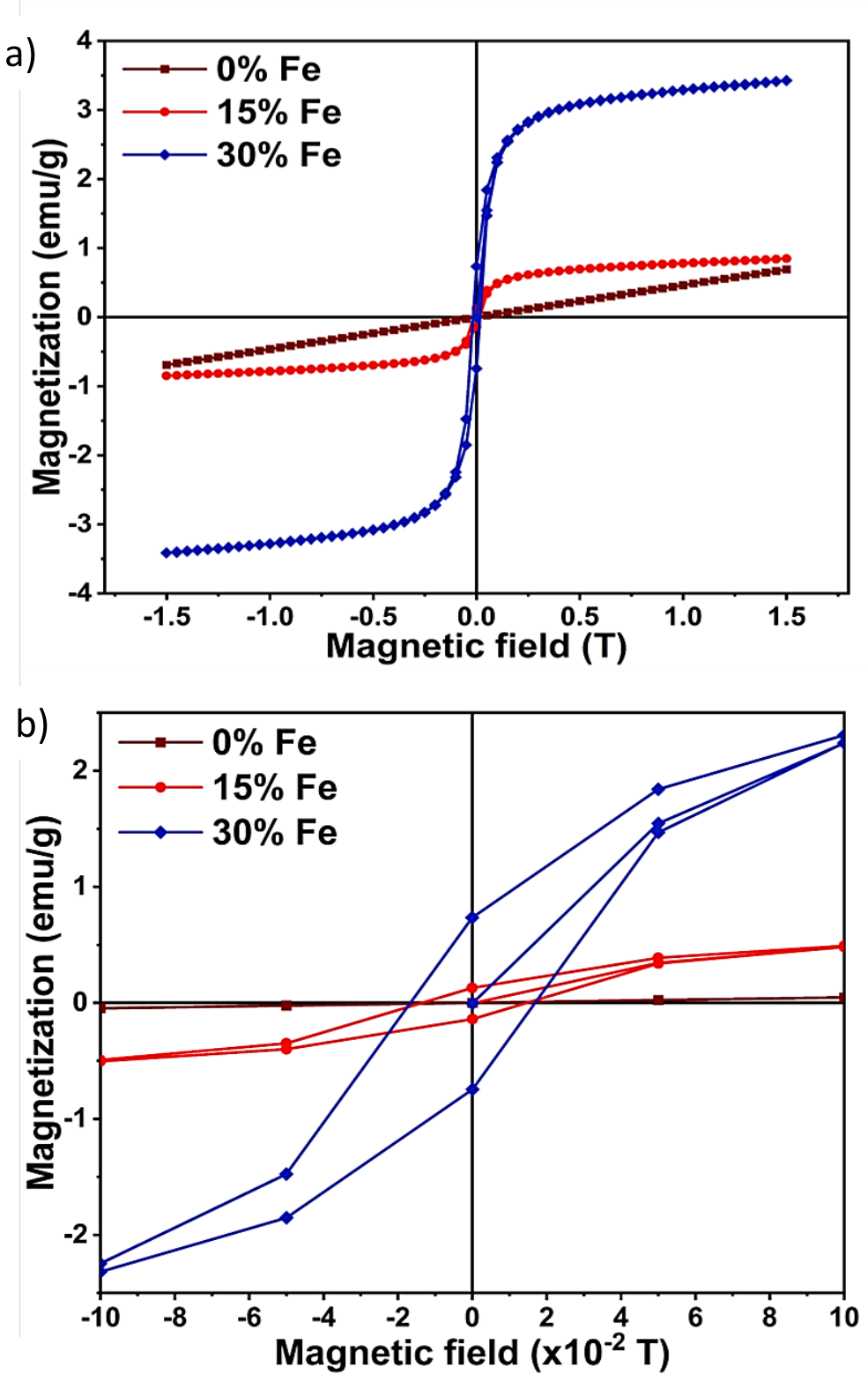}
		\caption{Room temperature VSM hysteresis curve of Fe-15 and Fe-30 NYF, compared with the pristine NYF sample is shown in the figure. Fig.(b) shows the enlarged view of the curve from which the magnetic remanence and coercitvity are calculated.}
		\label{vsm}
\end{figure}

The magnetic properties of pristine, Fe-15, Fe-30 NYF upconverting particles are measured using a Vibrating Sample Magnetometer (Fig.\ref{vsm}). The pristine sample exhibit linear M-H curves showing paramagnetic behaviour. The Fe-NYF's are ferromagnetic as seen from the S-shaped curve. The magnetization of 15\%Fe doped sample is $\approx$1 Am$^2$/kg at 1.5 T; however this increases to $\approx$3.5 Am$^2$/kg (at 1.5 T) for Fe-30 NYF upconverting particles. This material is in the context of standard superparamagnetic particles routinely used in the community of magnetic tweezers, like Dynabeads M-280 \cite{magneticbeads}, which has a saturation magnetisation of 6 A m$^2$/kg. It should be noted that the magnetization, in our particles, predominantly arises from the Fe dopant effect in NYF upconverting particles. In the powder of the particles, there is a presence of small concentration (<1\%) of Fe$_2$O$_3$ phase regions which are predominantly unstructured and can easily be distinguished from the hexagonal shaped UCP. If the magnetization contribution arose from Fe$_2$O$_3$, the magnetization values would have been much lower than 0.01  Am$^2$/kg \cite{hematite1,hematite2}. The realization of ferromagnetic property in Fe-NYF upconverting particles itself is interesting and not reported hitherto. The hysteresis curves of these magnetic particles also show discernible coercivity, i.e. ~ $\approx$20 mT. The magnetic properties of the particles with magnetization around few Am$^2$/kg and coercivity of $\approx$20 mT ideally suits the requirement whereby magnetic forces can be controlled by externally applied magnetic field.

\subsection*{Electronic structure calculations}

\begin{figure}[h]
	\centering
		\includegraphics[scale=0.12]{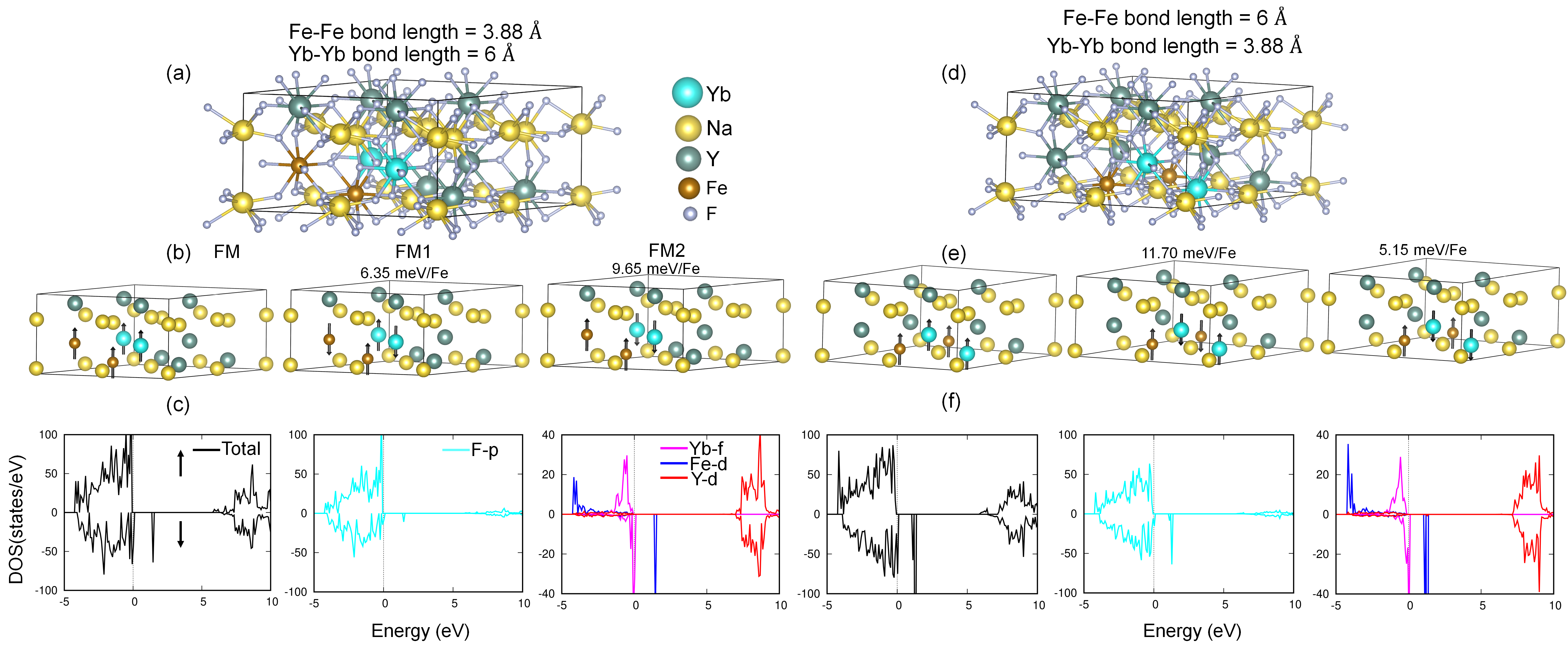}
		\caption{Top panel: Crystal structures of  NaY$_\frac{2}{3}$Yb$_\frac{1}{6}$Fe$_\frac{1}{6}$F$_4$ with Fe-Fe and Yb-Yb bond lengths being 3.88 \AA, 6 \AA and vice versa. Middle panel: The designed magnetic configurations: FM, FM1 and FM2. The energetics are shown with respect to the stable FM state. Bottom pannel: Spin and atom resolved total density of states of most stable FM configuration.}
		\label{dos}
\end{figure}

To gain more insights into the experimental observations of ferromagnetism in Fe-NYF, we examined the electronic structure by performing density functional theory (DFT) calculations within the framework of generalised gradient approximation (GGA) as implemented in Vienna
ab-initio simulation package (VASP). For this purpose we employed a supercell geometry and designed the composition NaY$_\frac{2}{3}$Yb$_\frac{1}{6}$Fe$_\frac{1}{6}$F$_4$. The Er doping is not taken into account as it has very low concentration in the experimentally synthesized particles. The supercell structures for the aforementioned composition with two Fe and two Yb atoms are shown in Fig. \ref{dos}. The structural and computational details are provided in the methods section. We have considered two doping arrangements. In one of them, the Fe-Fe (Yb-Yb) bond length is 3.88 (6) \AA  while in the another one it is 6 (3.88) \AA (see Figs. \ref{dos} (a,d)). For both arrangements, we have designed three magnetic configuration, namely, ferromagnet (FM), ferrimagnet1 (FM1) and ferrimagnet2 (FM2) (see middle panel of Fig. \ref{dos}). The DOS for the energetically most stable FM configuration are shown in the bottom panel of Fig. \ref{dos}.

The FM state is most stable as compared to the ferrimagnetic states for both doping arrangements and exhibits large ($\approx$ 4 $\mu_B$) and weak ($\approx$ 0.7 $\mu_B$) spin magnetic moments in Fe and Yb. The formation of large (weak) spin moment in Fe and Yb can be understood from the total density of states of Fe (Yb). Being the impurity states, the Fe-$d$ and Yb-$f$ states are well localized. For Fe, the majority spin-up channel is completely occupied whereas the minority spin-down channel is unoccupied giving rise to a large spin magnetic moment in Fe inferring a 3+ charge state which is also further confirmed from the XPS measurements. A detailed electronic structure analysis suggests that there is a reasonable hybridization among the Fe-$d$ and F-$p$ states leading to an induced magnetization in the latter. As a result, total magnetization per Fe is $\approx$ 5.1 $\mu_{B}$ with 4 $\mu_{B}$ contributed by the Fe-$d$ states and rest is contributed by the dispersive F-$p$ states. As Yb also possesses 3+ charge state and hence $4f^{13}$ electronic configuration, the down-spin channel is dominantly occupied and leads to the formation of a very weak local moment as compared to Fe. In the absence of Fe-doping, Yb demonstrates paramagnetism[]. However, with Fe-doping, the Yb spins develop ferromagnetic ordering with Fe to amplify the net magnetisation. With increased Fe-Fe bond length, the FM state remains energetically most favorable. Hence, the system exhibits a robust FM state which is in accordance with the experimental observations in this work. 

\subsection*{T-matrix calculation of trapping feasibility}
Optically trapping iron-oxide particles is known to be difficult, particularly if they are larger than a certain size. We investigate why that is the case and examine our Fe NYF particles towards suitability of trapping. The optical tweezers operation can be fundamentally treated as a scattering problem\cite{nieminen2007optical}. The particles of choice are frequently homogeneous and isotropic with spherical symmetry, for which an analytical solution to the scattering problem uses the generalized Lorentz-Mie theory (GLMT)\cite{gouesbet1990localized} or the transfer matrix approach. 

\begin{figure}[h]
	\centering
		\includegraphics[scale = 1.5]{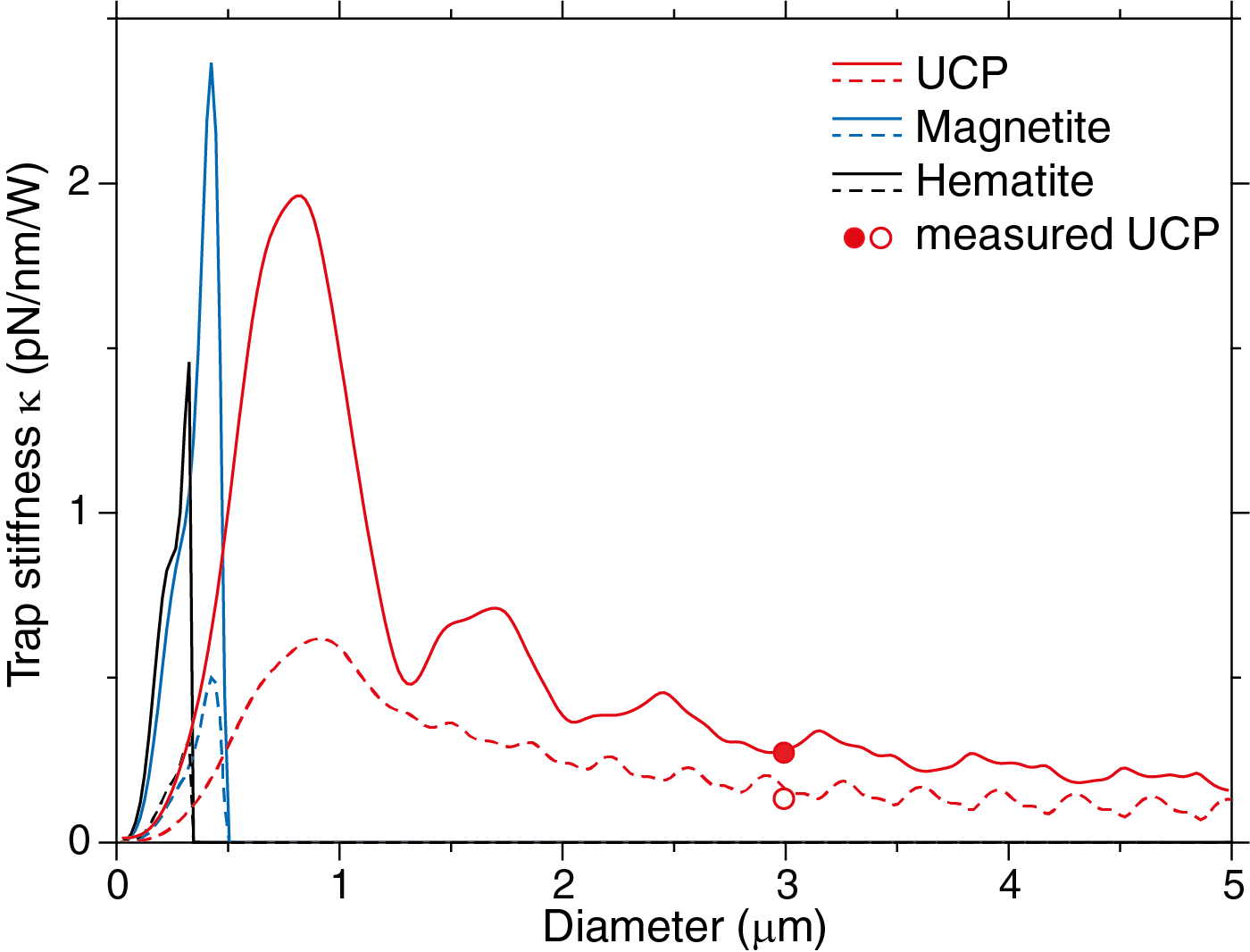}
	\caption{Lateral (solid lines) and axial (dashed lines) trap stiffness of magnetite, hematite and UCP particles as function of their diameter. The red solid (circle) data point represent the measured lateral (axial) trap stiffness of 3\,\textmu m UCP hexagonal particles. The scattering calculations are based on Mie-Theory. Here, the imaginary components of refractive index, implying absorption, has not been considered for Haematite and Magnetite. The trappable size reduces even further if the absorption is considered.}
	\label{OT_stiffness}
\end{figure}

We use this T-matrix method to perform optical trapping stability calculations using the optical tweezers toolbox \cite{nieminen2007optical, nieminen2007optical,nieminen2011t}.  We specify the parameters for the computation: refractive indices, $n_{UCP}=1.55$ \cite{refractiveindex}, $n_{hematite}=2.1 , n_{magnetite}=2.75$, at a laser wavelength of 1064\,nm, Numerical Aperture (NA) =1.3 and linearly polarized light. We calculate the lateral and axial trap stiffness as function of the diameter (Fig.~\ref{OT_stiffness}), assuming spherical particles. We find that due to their high refractive indices, magnetic Fe-oxide particles, like magnetite and hematite, are only stably trappable till a small diameter (up to 400\,nm). It may be noted here that the imaginary components of the refractive index (absorption) of haematite and magnetite have been ignored for this calculation. If these are included, the stable trapping diameter reduces even further. In contrast, Fe-doped upconverting particles have a low refractive index comparable to silica which allows them to be stably trapped in all three dimensions even for diameters of several micrometers. For example, 3\,\textmu m diameter spherical NYF's have a theoretical lateral (axial) trap stiffness of 0.36\,pN/nm/W (0.15\,pN/nm/W) that is in good agreement with the experimentally measured lateral (axial) trap stiffness of 0.26 $\pm$ 0.03 \,pN/nm/W (0.12$\pm$0.011\,pN/nm/W; mean$\pm$SD, $N$=10).

It can be noted here that the optical trapping of such Fe-NYF is unaffected by the absorption due to the Fe atoms. This possibly has to do with the Fe atoms being dopants in the material, implying low concentration. Thus, these Fe-NYF can be trapped without heating problems. 

\subsection*{Magneto-optical tweezers}

\begin{figure}[h]
	\centering
		\includegraphics[width=0.6\linewidth]{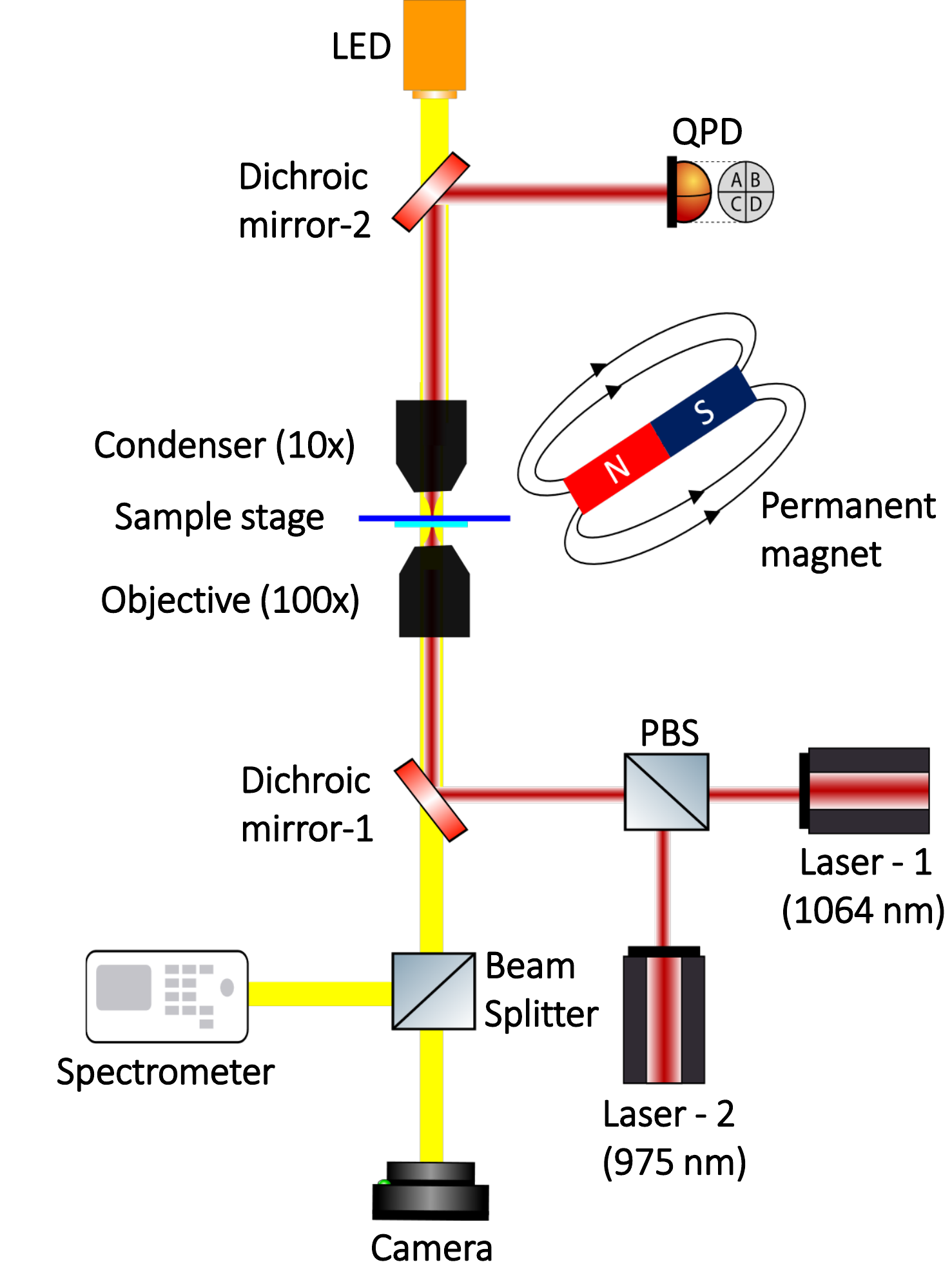}
	\caption{This figure represents the schematic diagram of the experiment. The entire setup consists of a single beam optical tweezers configured as an inverted microscope and its detection units. The optical trap is created near the focus of the 100x, 1.3 N.A objective and the entire sample chamber is illuminated by an LED through a dichroic mirror as shown in the figure.}
	\label{schematic}
\end{figure}

One such Fe-15 NYF upconverting particle was optically trapped and then a magnet moved in the proximity as indicated in Fig.~\ref{schematic}. 

\begin{figure}
	\centering
		\includegraphics[width=0.8\linewidth]{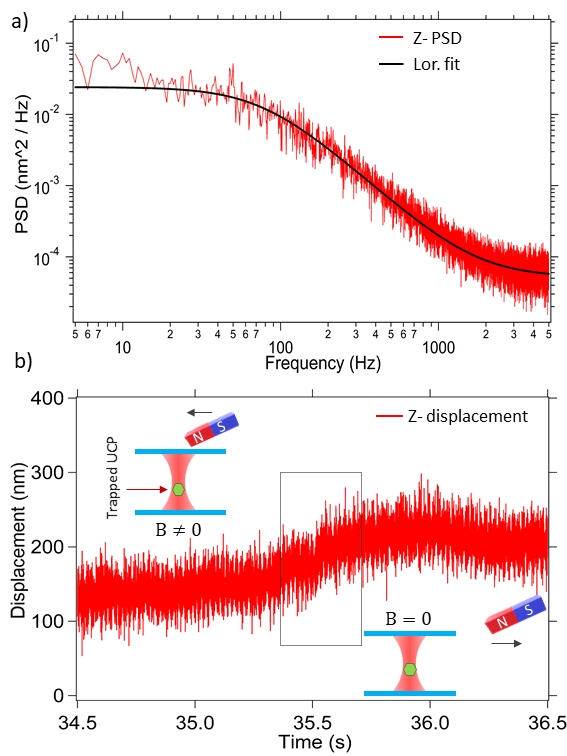}
	\caption{The power spectral density of perpendicular displacements (z-displacements) of an optically trapped Fe-15 NYF is shown in Fig (a). The optical trap is calibrated with a displacement sensitivity of 1.0 $\pm$ 0.1$\mu$m / Volt. The corner frequency ($f_c$) is 103 $\pm$ 10 Hz. In (b), the calibrated perpendicular displacement(z) time series of a Fe-15 NYF in the optical trap, exposed to an external magnetic field (B) generated by an electromagnet is shown. This roughly corresponds to a 6 $\pm$ 1 pN force by the magnetic field. The marked region in the time series indicates the time interval during which the field, B is turned off. The displacements are recorded using photonic force microscopy with more sensitivity than that available with video tracking for the axial direction motion.  }
	\label{timeseries}
\end{figure}

The power spectral density for an optically trapped Fe-15NYF at 1064\,nm wavelength has been shown in the Fig.~\ref{timeseries}(a) and the subsequent motion of the particle in the z direction after the magnetic field is removed is shown in Fig. \ref{timeseries}(b). It may be noted here that the axial position of the particle has been detected to a bandwidth of 10 kHz using the interferometric detection made possible by the optical tweezers, something difficult to do with video tracking at high resolution.

\begin{figure}[h]
	\centering
		\includegraphics[width=0.6\linewidth]{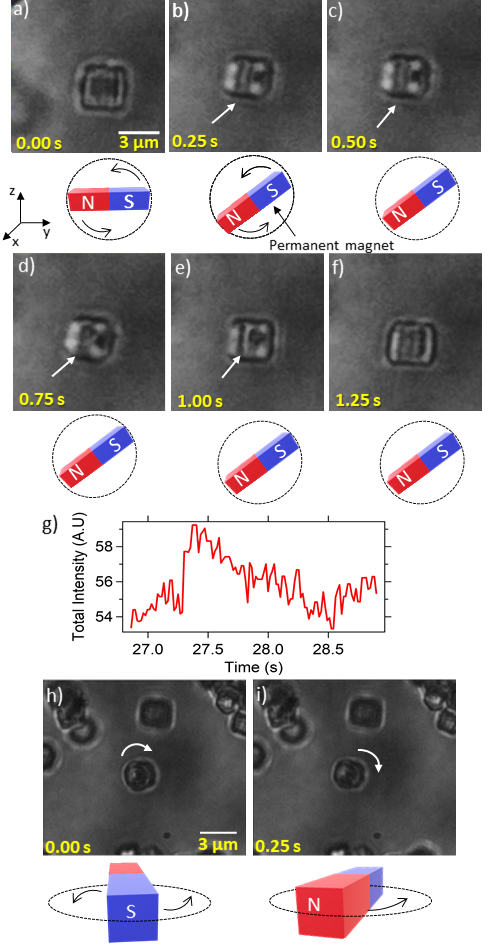}
	\caption{Figures (a-f) show the pitch rotational motion of an optically trapped, Fe-15 NYF upconverting particle under the influence of an external magnetic field. Fig (g) represents the corresponding time series of rotation, where the total intensity inside a rectangular region located at the tip of the arrow in figure (b) has been considered. In figures (h-i), we also show the yaw rotations of the particle under the influence of same external magnetic field.}
	\label{pitch}
\end{figure}

\begin{figure}[h]
	\centering
		\includegraphics[width=0.6\linewidth]{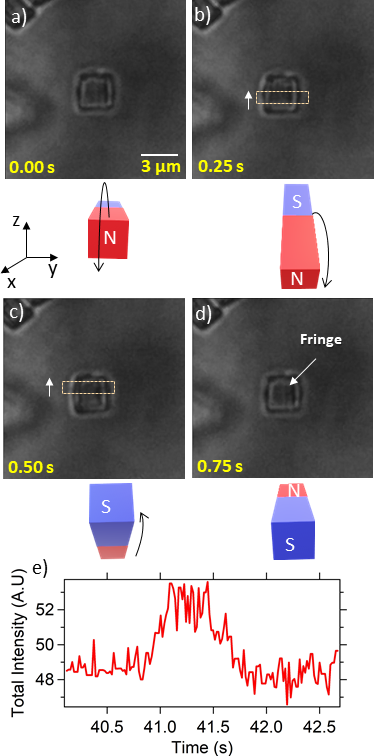}
	\caption{In figures (a-d), the roll rotational motion of an optically trapped Fe-15 NYF upconverting particle under the influence of an external magnetic field is shown. And the time series (e) corresponding to the roll motion of the total intensity inside one of the rectangular regions is also shown. The marked fringe on the particle is taken as a reference to visualize the out-of-plane roll motion. Arrows near the bar magnet indicates the direction in which it is rotated to generate roll motion.}
	\label{roll}
\end{figure}

We find that the particle turns in all rotational senses. First we turn in the pitch sense as indicated in Fig.~\ref{pitch}. The out of plane motion is visible. The region indicating the out of plane motion was then selected in the image sequence and a time series recorded to ascertain the dynamics (See Fig.~\ref{pitch}(g)). Then we turn the particle in the in-plane yaw sense, as indicated in the Fig.~\ref{pitch}(h) and (i). 
Then, we direct our attention to turn the particle in the roll sense as indicated in the Fig, \ref{roll}. Here, like the pitch motion, the region of the image indicating the roll motion was selected and a time series of the total intensity in that region recorded to find the dynamics. In this way, the roll motion can also be detected using the camera.  

\subsection*{Heat generation with NYF}

\begin{figure}[h]
	\centering
		\includegraphics[width=0.8\linewidth]{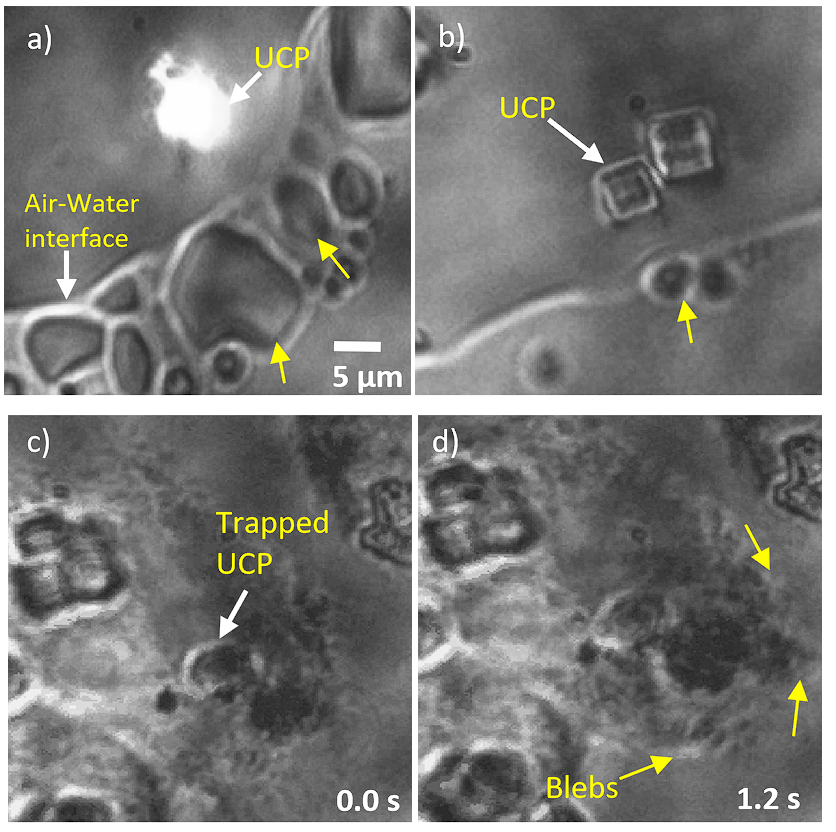}
	\caption{The initiation of boiling of water to form vapour bubbles, at the air-water interface due to the heating of an Fe-15 NYF particle, illuminated with (a) 975\,nm laser at 11.4\,mW power (where the visible emission from the particle appears as a white spot on the image, with spectrum in Fig. \ref{sem}(h)) and (b)  1064\,nm laser at 55\,mW power are shown. In figures (c-d), the blebbing of MCF-7 cells caused by the heating of an optically trapped Fe-15 NYF upconverting particle inside the cell is shown. The particle is trapped with 1064 nm laser at a power of 100 mW.}
	\label{blebb}
\end{figure}

We then go on to measure the thermal absorption properties of the particle. For this, we take one such particle and place it close to the side interface of a sessile water droplet on the glass cover slip. The particle is then illuminated with 1064\,nm light and 975\,nm light sequentially. We look for the laser power at which it starts to form bubbles on the side interface. We find that the threshold for bubbling on the side interface for 1064\,nm light is 55\,mW at the sample plane, while the same threshold power at 975\,nm is 11.4\,mW. Assuming that the formation of the bubbles are a consequence of the local environment of the droplet reaching the boiling point of water, we find that the NYF is 5 times less absorptive at 1064\,nm than at 975\,nm. This effect has been shown in Fig.~\ref{blebb} (a) and (b). To ascertain that the particle is responsible for absorption of the light and subsequent heating, we illuminate the 975\,nm laser beam without the particle and find no bubbles at the surface. Thus, the particle is indeed responsible for absorption and subsequent heating. It may be noted here that water itself is 3 times more absorptive at 975 nm than 1064 nm, which might partially account for increase in heating efficiency. It may also be noted here that the Fig. \ref{blebb}(a) shows a white patch which is the visible emission from the particle upon being illuminated by the 975 nm light with a spectrum shown in Fig. \ref{sem}(h). The backscattered infra-red light does not appear in the camera due to the dichroic in the backscatter direction which filters out any wavelength higher than 750 nm while only allowing visible light to pass. 

We also show a typical application of the heating of the particle by the lasers. We make cancer cells ingest the particles by phagocytosis and then illuminate them with a laser. When we illuminate the particle with about 100\,mW of 1064\,nm laser power, the cells die in seconds, as indicated by the formation of blebs in Fig.\ref{blebb}(c) and (d). The effect of killing the cell is also seen upon usage of 975 nm laser at much lower power. This is the first time that such an application has been shown with upconverting particles. It was believed that thermometry with UCP shows "apparent self-heating". We show that the effect was not apparent but an actual heating of the local environment by the particle.  If we prefer to not kill the cells, we use about 30\,mW of 1064 nm laser power and trap the particles while keeping the cell alive for more than 5 minutes. This is enough time to use the magnetic fields to apply forces.

\section*{Discussions}

These particles show emission in visible wavelengths when illuminated with 975 nm light by multiphoton processes like in Fig. \ref{sem}(h).  The spectra for the different configurations of the particle are different (as shown in Fig. \ref{sem}(h) ) which enables the possibility of using the changes in the spectra to determine the pitch rotational motion to a high resolution by illuminating with 975\,nm light itself \cite{amrendraEPJ,rodriguez2016optical,haro2013optical}.

 These can also then be used to generate continuous rolling motion of a free Fe-NYF upconverting particle using a magnetic field rotating along the out-of-plane sense \cite{rollingwithmagnet} to make a microrobot. The instantaneous angle at which it is rolling can be determined from the change in its emission spectra\cite{rodriguez2016optical}. These magnetic particles may also be used as magnetic microrollers on blood artery walls (or any vessel walls),  where the surface locomotion properties of such microrollers are determined by the shape anisotropy of the particles\cite{bozuyuk2021shape} and the rotational motion can be detected with high resolution from emission spectra.
 
 Some applications might require smaller sized particles. Hexagonal shaped NYF smaller than 500 nm have been routinely prepared. However, the exact protocol to prepare Fe-NYF particles smaller than 500 nm particles needs to be optimised. Recently, there has also been a lot of interest in generating large forces to deform the cytoskeleton of cells, typically several nano-Newtons. We can use these particles with large magnetic moments to apply large forces on the cell cytoskeleton to reach the non-linear regime of the cytoskeleton while accurately determining the position of the particle. 

\section*{Conclusions}

To summarize and conclude, we prepared  Fe co-doped NaYF$_4$: Yb, Er upconversion particles that have ferromagnetic properties.  The upconversion properties of the particles are confirmed from their emission spectra. We find that a single Fe doped upconverting particle of diameter in excess of 3 $\mu$m is optically trappable in three-dimensions and the dynamics of such a trapped particle may be controlled in all six dimensions of a rigid body enabling multifarious uses. The optical trap parameters determined from our experiments are close to the results from scattering simulations performed using the T-matrix tool. By applying an external magnetic field, an optically trapped particle may be rotated in yaw, pitch, and roll sense. We detect the rotations with a video camera and the translations using a quadrant photo diode. We show that these particle get significantly heated upon absorption of infra-red light, in particular 975 nm pump wavelength, which might compare favorably with using gold nanoparticles for heating\cite{goldnanoheating}. We use this to kill cancer cells in photo-thermal therapy\cite{photothermaltherapy}, which can be made even more efficient by using alternating magnetic fields. The exact comparison of such NYF and gold towards efficiency of photothermal therapy is beyond the scope of this manuscript. There is also a possibility to use these particles as microrobots by rolling them on surfaces using rotating magnetic fields and detecting the rotation by the change in spectra with a very low power 975 nm light beam. These new particles can be moved to the desired location inside the body by magnetic and optical means, and then illuminated with 975 nm light to locally heat the environment and kill cancerous cells. This can have applications in combining the best of optical tweezers and magnetic forces to enable a new class of experiments where application of forces and torques with high-bandwidth accurate position and orientation sensitivity is required. 

\section*{Methods}
\section{Materials used}
RE(NO$_3$)$_3$.6H$_2$O (RE=Y,Yb and Er, 99.99\% purity), Fe(NO$_3$)$_3$.9H$_2$O (99.999\%) and sodium citrate dihydrate (99\% purity) were purchased from Alfa Aesar and were used without further purification. Double distilled water and C$_2$H$_5$OH (Hayman, 99.9\%) were used to prepare Fe-NYF upconverting particles.
\section{Preparation method}

The Fe-doped NaYF$_4$:Yb,Er upconverting particles (Fe-NYF) were prepared by the basic hydrothermal method with some modifications \cite{kumar2020pitch}. However, citrate assisted hydrothermal preparation of Fe-NYF upconverting particles has not yet been reported. To prepare Fe-15 NYF, 0.126\,M of Y(NO$_3$)$_3$ (63 at.\% of Y) and 0.3\,M of sodium citrate were added to 14\, mL of water with vigorous magnetic stirring for 10\,min. Then, 40\,mM Yb(NO$_3$)$_3$ (20 at.\% Yb), 4 mM Er(NO$_3$)$_3$ (2 at.\% Er) and 30 mM Fe(NO$_3$)$_3$ (15 at.\% Fe) in 21\,mL of H$_2$O (taken from Millipore DQ 3, Merck Systems) were added into the sodium citrate solution, which then appeared as a milky white solution. A transparent solution was obtained by adding 0.5 M of NaF in 67mL of aqueous solution into the milky white solution and stirred for 1\,hour. The transparent solution  was subsequently transferred to 200\,mL Teflon lined autoclave and tightly sealed. Then, it was heated using muffle furnace at 200\,$^{\circ}$C for 12\,hours. The solution was washed with water and ethanol four times after cooling naturally to room temperature inside the autoclave. The sample in powdery form was collected after drying at 100\,$^{\circ}$C for 12 hours. Similar preparation steps were also followed to prepare Fe-30 NYF except change in the concentration to 0.06\,M of Fe(NO$_3$)$_3$ (30 at.\% Fe) and 0.096\,M of Y(NO$_3$)$_3$ (48 at.\% of Y). These particles were dispersed in de-ionized water for the experiment.

\section{Experimental details}

The experiments are performed on a setup made from optical tweezers kit OTKB/M (Thorlabs) in the inverted configuration. Either a linearly polarised laser beam (Diode laser :  LSR1064NL, Lasever, China) of wavelength 1064\,nm and maximum output power 1.5\,W  or a a linearly polarized 400\, mW, 975\,nm butterfly laser (CLD1015, Thorlabs) is used to form the optical trap. The laser is tightly focused on the sample chamber using an Olympus 100X objective (1.3 NA, oil immersion type) and a 10X Nikon condenser (0.25 NA, air immersion) is used to collect the forward scattered light from the optical trap as shown in Fig.\ref{schematic}. The power at the sample stage of the 1064 nm laser is 65\,mW, measured with a power meter (PM100D, Thorlabs) with a head unit S130C (Thorlabs). An LED lamp is used to illuminate the sample from the top via a dichroic mirror, as shown in Fig.~\ref{schematic}. The forward scattered light is made incident on a quadrant photo diode (QPD) (PDQ80A, Thorlabs, USA) upon reflection from the same dichroic mirror.

The sample chamber consists of a glass slide of size 75\,mm\ $\times$ 25\,mm\ $\times$ 1\,mm (Blue star, number 1 size, English glass) on the top of which the sample is placed using a pipette and mounted with a glass cover slip (Blue star, number 1 size, English glass). The whole system is then inverted and put on the sample stage. A piezo-electric stage (Thorlabs Nanomax 300) is used to hold the sample in between the objective and condenser lenses. A linearly polarized laser beam of wavelength 1064\,nm is tightly focused on a single UCP which constitutes the optical tweezers. Water was taken from taken from Millipore (DQ 3, Merck Systems). 

These particles tend to aggregate when left unattended.  There are some Fe$_2$O$_3$ regions in the sample which are clearly distinguishable, being of an unstructured morphology than the regular shaped hexagonal particles. When these regions are illuminated with infra-red light, we tend to get heating due to absorption and then the aggregate explodes. This is how we get individual hexagonal particles. 

These magnetic particles are manipulated manually using a ring type magnet (Ceramic ferrite magnet, Perfect magnet, India) of inner radius 22\,mm, outer radius 45\,mm and thickness 20\,mm. A magnetic field of 30 Gauss is produced by this magnet at an axial distance of 8\,cm. The magnet is rotated manually with respect to X and Y axes of the piezo stage to produce yaw, pitch and roll motions of the trapped particle. The response of an optically trapped magnetic upconverting particle to the external field is recorded using a CMOS camera (CS165CU1, Thorlabs). The particles we trapped about 10 $\mu$m from the bottom surface. The entire system was kept at room temperature (25\,$^{\circ}$C).

For the air-water interface experiment, 2\,$\mu$L of the sample is placed on the glass slide (Blue star, number 1 size, English glass) and mounted over the sample stage.  The two laser beams are alternately used to illuminate the sample. The heating effects at the interface due to the particle are recorded by turning on one beam at a time and varying its power.

10\,$\mu$L of 15\% Fe doped upconverting particles is added to MCF-7 cells (obtained from National Center for Cell Science, Pune, India and maintained in Dulbecco’s Modified Eagle’s Media) and incubated overnight to allow phagocytosis to take place. The cells are washed with 1X PBS to remove excess particles and fresh media are added. One such 15\% Fe doped particle is trapped inside the cell, first with 1064\,nm and latter with 975\,nm laser beams as shown in Fig.~\ref{schematic}.

\section{Structural and Computational Details}

We have designed NaY$_\frac{2}{3}$Yb$_\frac{1}{6}$Fe$_\frac{1}{6}$F$_4$ configuration to theoretically understand the origin of ferromagnetism as observed experimentally. In this configuration, Er doping is not taken into account as it has very low concentration and will have negligible effect on the electronic and magnetic structure of the system. The NaY$_\frac{2}{3}$Yb$_\frac{1}{6}$Fe$_\frac{1}{6}$F$_4$ configuration is designed by performing the following steps: (i) Na$_4$Y$_2$F$_6$ crystal structure is constructed with P6$_3$/m space group symmetry where Na, Y and F atoms occupies the $4e$, $2d$ and $6h$ wyckoff sites, respectively. The $4e$ and $2d$ wyckoff sites are one-fourth and three-fourth occupied whereas the $6h$ site is completely occupied. (ii) The obtained crystal structure is doubled along the c-axis. Further, since the Na-$4e$ site is one-fourth occupied, six Na atoms are removed from the structure. One of the Y atoms is replaced by Na as the $2d$ site is shared by Y and Na atoms with three-fourth and one-fourth occupancy. The resultant crystal structure is Na$_3$Y$_3$F$_{12}$. Further, to design NaY$_\frac{2}{3}$Yb$_\frac{1}{6}$Fe$_\frac{1}{6}$F$_4$ configuration, a 2 $\times$ 2 $\times$ 1 supercell is designed out of Na$_3$Y$_3$F$_{12}$ and four of the Y sites are replaced by two Fe and two Yb atoms (see Figs. \ref{dos}(a,d)).\par
To find the plausible origin of ferromagnetism in the system, the first-principles based density functional theory (DFT) calculations were carried out in the plane-wave basis using the projector augmented wave potentials \cite{Kresse1999} as implemented in Vienna ab-initio simulation package (VASP)\cite{Kresse1996}.
The generalized gradient approximation (GGA) was chosen for the exchange-correlation functional. The Brillouin zone integration were carried out using a 4 $\times$ 4 $\times$ 8 and 8 $\times$ 8 $\times$ 8 Gamma-centered k-mesh for achieving the self-consistent ground state and its corresponding density of states. A 400\,eV kinetic energy cutoff was chosen for the plane wave basis set.\par

\section*{Acknowledgements}
We thank the Indian Institute of Technology, Madras, India for their seed and initiation grants. This work was also supported by the DBT/Wellcome Trust India Alliance Fellowship IA/I/20/1/504900 awarded to Basudev Roy. We would like to thank HPCE, IIT Madras for providing the computational facility. This work is also funded by the Department of Science and Technology, India, through grant No. CRG/2020/004330. We also thank Privita Edwina and Saumendra Bajpai for providing MCF7 cells and 
putting UCP inside.

\section*{Author contributions statement}

G.N, G.M, M.L and R.V performed the experiment and analysed data. G.M and J.S synthesized the particles. A.C and B.R.K.N performed the density functional theory calculations. C.S analysed TEM and XPS data. A.J and E.S analysed data and performed the scattering calculations. B.R conceptualised the experiment and analysed data. B.R, C.S, A.C, B.R.K.N and E.S Wrote the manuscript.
\bibliography{cas-refs}

\begin{thebibliography}{10}
\urlstyle{rm}
\expandafter\ifx\csname url\endcsname\relax
  \def\url#1{\texttt{#1}}\fi
\expandafter\ifx\csname urlprefix\endcsname\relax\def\urlprefix{URL }\fi
\expandafter\ifx\csname doiprefix\endcsname\relax\def\doiprefix{DOI: }\fi
\providecommand{\bibinfo}[2]{#2}
\providecommand{\eprint}[2][]{\url{#2}}

\bibitem{joo2008advances}
\bibinfo{author}{Joo, C.}, \bibinfo{author}{Balci, H.},
  \bibinfo{author}{Ishitsuka, Y.}, \bibinfo{author}{Buranachai, C.} \&
  \bibinfo{author}{Ha, T.}
\newblock \bibinfo{journal}{\bibinfo{title}{Advances in single-molecule
  fluorescence methods for molecular biology}}.
\newblock {\emph{\JournalTitle{Annu. Rev. Biochem.}}}
  \textbf{\bibinfo{volume}{77}}, \bibinfo{pages}{51--76}
  (\bibinfo{year}{2008}).

\bibitem{holden2010defining}
\bibinfo{author}{Holden, S.~J.} \emph{et~al.}
\newblock \bibinfo{journal}{\bibinfo{title}{Defining the limits of
  single-molecule fret resolution in tirf microscopy}}.
\newblock {\emph{\JournalTitle{Biophys. J}}} \textbf{\bibinfo{volume}{99}},
  \bibinfo{pages}{3102--3111} (\bibinfo{year}{2010}).

\bibitem{cheng2011single}
\bibinfo{author}{Cheng, W.}, \bibinfo{author}{Arunajadai, S.~G.},
  \bibinfo{author}{Moffitt, J.~R.}, \bibinfo{author}{Tinoco, I.} \&
  \bibinfo{author}{Bustamante, C.}
\newblock \bibinfo{journal}{\bibinfo{title}{Single--base pair unwinding and
  asynchronous rna release by the hepatitis c virus ns3 helicase}}.
\newblock {\emph{\JournalTitle{Science}}} \textbf{\bibinfo{volume}{333}},
  \bibinfo{pages}{1746--1749} (\bibinfo{year}{2011}).

\bibitem{abbondanzieri2005direct}
\bibinfo{author}{Abbondanzieri, E.~A.}, \bibinfo{author}{Greenleaf, W.~J.},
  \bibinfo{author}{Shaevitz, J.~W.}, \bibinfo{author}{Landick, R.} \&
  \bibinfo{author}{Block, S.~M.}
\newblock \bibinfo{journal}{\bibinfo{title}{Direct observation of base-pair
  stepping by rna polymerase}}.
\newblock {\emph{\JournalTitle{Nature}}} \textbf{\bibinfo{volume}{438}},
  \bibinfo{pages}{460--465} (\bibinfo{year}{2005}).

\bibitem{greenleaf2005passive}
\bibinfo{author}{Greenleaf, W.~J.}, \bibinfo{author}{Woodside, M.~T.},
  \bibinfo{author}{Abbondanzieri, E.~A.} \& \bibinfo{author}{Block, S.~M.}
\newblock \bibinfo{journal}{\bibinfo{title}{Passive all-optical force clamp for
  high-resolution laser trapping}}.
\newblock {\emph{\JournalTitle{Phys. Rev. Lett.}}}
  \textbf{\bibinfo{volume}{95}}, \bibinfo{pages}{208102}
  (\bibinfo{year}{2005}).

\bibitem{koster2005friction}
\bibinfo{author}{Koster, D.~A.}, \bibinfo{author}{Croquette, V.},
  \bibinfo{author}{Dekker, C.}, \bibinfo{author}{Shuman, S.} \&
  \bibinfo{author}{Dekker, N.~H.}
\newblock \bibinfo{journal}{\bibinfo{title}{Friction and torque govern the
  relaxation of dna supercoils by eukaryotic topoisomerase ib}}.
\newblock {\emph{\JournalTitle{Nature}}} \textbf{\bibinfo{volume}{434}},
  \bibinfo{pages}{671--674} (\bibinfo{year}{2005}).

\bibitem{strick2000single}
\bibinfo{author}{Strick, T.~R.}, \bibinfo{author}{Croquette, V.} \&
  \bibinfo{author}{Bensimon, D.}
\newblock \bibinfo{journal}{\bibinfo{title}{Single-molecule analysis of dna
  uncoiling by a type ii topoisomerase}}.
\newblock {\emph{\JournalTitle{Nature}}} \textbf{\bibinfo{volume}{404}},
  \bibinfo{pages}{901--904} (\bibinfo{year}{2000}).

\bibitem{tirf1}
\bibinfo{author}{Fish, K.~N.}
\newblock \bibinfo{journal}{\bibinfo{title}{Total internal reflection
  fluorescence (tirf) microscopy}}.
\newblock {\emph{\JournalTitle{Current protocols in cytometry}}}
  \textbf{\bibinfo{volume}{50}}, \bibinfo{pages}{12--18}
  (\bibinfo{year}{2009}).

\bibitem{tirf2}
\bibinfo{author}{Axelrod, D.}
\newblock \bibinfo{journal}{\bibinfo{title}{Total internal reflection
  fluorescence microscopy in cell biology}}.
\newblock {\emph{\JournalTitle{Traffic}}} \textbf{\bibinfo{volume}{2}},
  \bibinfo{pages}{764--774} (\bibinfo{year}{2001}).

\bibitem{dulin2013studying}
\bibinfo{author}{Dulin, D.}, \bibinfo{author}{Lipfert, J.},
  \bibinfo{author}{Moolman, M.~C.} \& \bibinfo{author}{Dekker, N.~H.}
\newblock \bibinfo{journal}{\bibinfo{title}{Studying genomic processes at the
  single-molecule level: introducing the tools and applications}}.
\newblock {\emph{\JournalTitle{Nat. Rev. Genet.}}}
  \textbf{\bibinfo{volume}{14}}, \bibinfo{pages}{9--22} (\bibinfo{year}{2013}).

\bibitem{kimura2006single}
\bibinfo{author}{Kimura, Y.} \& \bibinfo{author}{Bianco, P.~R.}
\newblock \bibinfo{journal}{\bibinfo{title}{Single molecule studies of dna
  binding proteins using optical tweezers}}.
\newblock {\emph{\JournalTitle{Analyst}}} \textbf{\bibinfo{volume}{f131}},
  \bibinfo{pages}{868--874} (\bibinfo{year}{2006}).

\bibitem{neuman2004optical}
\bibinfo{author}{Neuman, K.~C.} \& \bibinfo{author}{Block, S.~M.}
\newblock \bibinfo{journal}{\bibinfo{title}{Optical trapping}}.
\newblock {\emph{\JournalTitle{Rev. Sci. Instrum.}}}
  \textbf{\bibinfo{volume}{75}}, \bibinfo{pages}{2787--2809}
  (\bibinfo{year}{2004}).

\bibitem{jannasch2012nanonewton}
\bibinfo{author}{Jannasch, A.}, \bibinfo{author}{Demir{\"o}rs, A.~F.},
  \bibinfo{author}{Van~Oostrum, P.~D.}, \bibinfo{author}{Van~Blaaderen, A.} \&
  \bibinfo{author}{Sch{\"a}ffer, E.}
\newblock \bibinfo{journal}{\bibinfo{title}{Nanonewton optical force trap
  employing anti-reflection coated, high-refractive-index titania
  microspheres}}.
\newblock {\emph{\JournalTitle{Nat. Photonics}}} \textbf{\bibinfo{volume}{6}},
  \bibinfo{pages}{469--473} (\bibinfo{year}{2012}).

\bibitem{ref10}
\bibinfo{author}{Conroy, R.}
\newblock \bibinfo{title}{Force spectroscopy with optical and magnetic
  tweezers}.
\newblock In \emph{\bibinfo{booktitle}{Handbook of molecular force
  spectroscopy}}, \bibinfo{pages}{Springer: New York, NY, USA \, 23--96}
  (\bibinfo{year}{2008}).

\bibitem{ref11}
\bibinfo{author}{Strick, T.~R.}, \bibinfo{author}{Allemand, J.-F.},
  \bibinfo{author}{Bensimon, D.}, \bibinfo{author}{Bensimon, A.} \&
  \bibinfo{author}{Croquette, V.}
\newblock \bibinfo{journal}{\bibinfo{title}{The elasticity of a single
  supercoiled dna molecule}}.
\newblock {\emph{\JournalTitle{Science}}} \textbf{\bibinfo{volume}{271}},
  \bibinfo{pages}{1835--1837} (\bibinfo{year}{1996}).

\bibitem{ref12}
\bibinfo{author}{Gosse, C.} \& \bibinfo{author}{Croquette, V.}
\newblock \bibinfo{journal}{\bibinfo{title}{Magnetic tweezers:
  micromanipulation and force measurement at the molecular level}}.
\newblock {\emph{\JournalTitle{Biophys. J.}}} \textbf{\bibinfo{volume}{82}},
  \bibinfo{pages}{3314--3329} (\bibinfo{year}{2002}).

\bibitem{bausch1999measurement}
\bibinfo{author}{Bausch, A.~R.}, \bibinfo{author}{M{\"o}ller, W.} \&
  \bibinfo{author}{Sackmann, E.}
\newblock \bibinfo{journal}{\bibinfo{title}{Measurement of local
  viscoelasticity and forces in living cells by magnetic tweezers}}.
\newblock {\emph{\JournalTitle{Biophys. J}}} \textbf{\bibinfo{volume}{76}},
  \bibinfo{pages}{573--579} (\bibinfo{year}{1999}).

\bibitem{neuman2008single}
\bibinfo{author}{Neuman, K.~C.} \& \bibinfo{author}{Nagy, A.}
\newblock \bibinfo{journal}{\bibinfo{title}{Single-molecule force spectroscopy:
  optical tweezers, magnetic tweezers and atomic force microscopy}}.
\newblock {\emph{\JournalTitle{Nat. methods}}} \textbf{\bibinfo{volume}{5}},
  \bibinfo{pages}{491--505} (\bibinfo{year}{2008}).

\bibitem{van2012non}
\bibinfo{author}{van Loenhout, M.~T.}, \bibinfo{author}{Kerssemakers, J.~W.},
  \bibinfo{author}{De~Vlaminck, I.} \& \bibinfo{author}{Dekker, C.}
\newblock \bibinfo{journal}{\bibinfo{title}{Non-bias-limited tracking of
  spherical particles, enabling nanometer resolution at low magnification}}.
\newblock {\emph{\JournalTitle{Biophys. J}}} \textbf{\bibinfo{volume}{102}},
  \bibinfo{pages}{2362--2371} (\bibinfo{year}{2012}).

\bibitem{cnossen2014optimized}
\bibinfo{author}{Cnossen, J.~P.}, \bibinfo{author}{Dulin, D.} \&
  \bibinfo{author}{Dekker, N.}
\newblock \bibinfo{journal}{\bibinfo{title}{An optimized software framework for
  real-time, high-throughput tracking of spherical beads}}.
\newblock {\emph{\JournalTitle{Rev. Sci. Instrum.}}}
  \textbf{\bibinfo{volume}{85}}, \bibinfo{pages}{103712}
  (\bibinfo{year}{2014}).

\bibitem{gosse2002magnetic}
\bibinfo{author}{Gosse, C.} \& \bibinfo{author}{Croquette, V.}
\newblock \bibinfo{journal}{\bibinfo{title}{Magnetic tweezers:
  micromanipulation and force measurement at the molecular level}}.
\newblock {\emph{\JournalTitle{Biophys. J.}}} \textbf{\bibinfo{volume}{82}},
  \bibinfo{pages}{3314--3329} (\bibinfo{year}{2002}).

\bibitem{lansdorp2013high}
\bibinfo{author}{Lansdorp, B.~M.}, \bibinfo{author}{Tabrizi, S.~J.},
  \bibinfo{author}{Dittmore, A.} \& \bibinfo{author}{Saleh, O.~A.}
\newblock \bibinfo{journal}{\bibinfo{title}{A high-speed magnetic tweezer
  beyond 10,000 frames per second}}.
\newblock {\emph{\JournalTitle{Rev. Sci. Instrum.}}}
  \textbf{\bibinfo{volume}{84}}, \bibinfo{pages}{044301}
  (\bibinfo{year}{2013}).

\bibitem{huhle2015camera}
\bibinfo{author}{Huhle, A.} \emph{et~al.}
\newblock \bibinfo{journal}{\bibinfo{title}{Camera-based three-dimensional
  real-time particle tracking at khz rates and {\aa}ngstr{\"o}m accuracy}}.
\newblock {\emph{\JournalTitle{Nat. Commun.}}} \textbf{\bibinfo{volume}{6}},
  \bibinfo{pages}{1--8} (\bibinfo{year}{2015}).

\bibitem{photonics2030758}
\bibinfo{author}{Zhou, Z.}, \bibinfo{author}{Miller, H.},
  \bibinfo{author}{Wollman, A.~J.} \& \bibinfo{author}{Leake, M.~C.}
\newblock \bibinfo{journal}{\bibinfo{title}{Developing a new biophysical tool
  to combine magneto-optical tweezers with super-resolution fluorescence
  microscopy}}.
\newblock {\emph{\JournalTitle{Photonics}}} \textbf{\bibinfo{volume}{2}},
  \bibinfo{pages}{758--772} (\bibinfo{year}{2015}).

\bibitem{crut2007fast}
\bibinfo{author}{Crut, A.}, \bibinfo{author}{Koster, D.~A.},
  \bibinfo{author}{Seidel, R.}, \bibinfo{author}{Wiggins, C.~H.} \&
  \bibinfo{author}{Dekker, N.~H.}
\newblock \bibinfo{journal}{\bibinfo{title}{Fast dynamics of supercoiled dna
  revealed by single-molecule experiments}}.
\newblock {\emph{\JournalTitle{Proc.Natl. Acad. Sci. U.S.A}}}
  \textbf{\bibinfo{volume}{104}}, \bibinfo{pages}{11957--11962}
  (\bibinfo{year}{2007}).

\bibitem{sumitfrontinphys}
\bibinfo{author}{Kumar, S.} \emph{et~al.}
\newblock \bibinfo{journal}{\bibinfo{title}{Trapped in out-of-equilibrium
  stationary state: Hot brownian motion in optically trapped upconverting
  nanoparticles}}.
\newblock {\emph{\JournalTitle{Front. Phys.}}} \textbf{\bibinfo{volume}{8}},
  \bibinfo{pages}{429} (\bibinfo{year}{2020}).

\bibitem{kumar2020pitch}
\bibinfo{author}{Kumar, S.} \emph{et~al.}
\newblock \bibinfo{journal}{\bibinfo{title}{Pitch-rotational manipulation of
  single cells and particles using single-beam thermo-optical tweezers}}.
\newblock {\emph{\JournalTitle{Biomed. Opt. Express.}}}
  \textbf{\bibinfo{volume}{11}}, \bibinfo{pages}{3555--3566}
  (\bibinfo{year}{2020}).

\bibitem{sumitnanoscale}
\bibinfo{author}{Kumar, S.}, \bibinfo{author}{Gunaseelan, M.},
  \bibinfo{author}{Vaippully, R.}, \bibinfo{author}{Banerjee, A.} \&
  \bibinfo{author}{Roy, B.}
\newblock \bibinfo{journal}{\bibinfo{title}{Breaking the diffraction limit in
  absorption spectroscopy using upconverting nanoparticles}}.
\newblock {\emph{\JournalTitle{Nanoscale}}} \textbf{\bibinfo{volume}{13}},
  \bibinfo{pages}{11856--11866} (\bibinfo{year}{2021}).

\bibitem{friese1998optical}
\bibinfo{author}{Friese, M.~E.}, \bibinfo{author}{Nieminen, T.~A.},
  \bibinfo{author}{Heckenberg, N.~R.} \& \bibinfo{author}{Rubinsztein-Dunlop,
  H.}
\newblock \bibinfo{journal}{\bibinfo{title}{Optical alignment and spinning of
  laser-trapped microscopic particles}}.
\newblock {\emph{\JournalTitle{Nature}}} \textbf{\bibinfo{volume}{394}},
  \bibinfo{pages}{348--350} (\bibinfo{year}{1998}).

\bibitem{rahulyaw}
\bibinfo{author}{Vaippully, R.}, \bibinfo{author}{Bhatt, D.},
  \bibinfo{author}{Ranjan, A.~D.} \& \bibinfo{author}{Roy, B.}
\newblock \bibinfo{journal}{\bibinfo{title}{Study of adhesivity of surfaces
  using rotational optical tweezers}}.
\newblock {\emph{\JournalTitle{Phys. Scr.}}} \textbf{\bibinfo{volume}{94}},
  \bibinfo{pages}{105008} (\bibinfo{year}{2019}).

\bibitem{rahulsyaw2}
\bibinfo{author}{Vaippully, R.}, \bibinfo{author}{Lokesh, M.} \&
  \bibinfo{author}{Roy, B.}
\newblock \bibinfo{journal}{\bibinfo{title}{Continuous rotational motion in
  birefringent particles using two near-orthogonally polarized optical tweezers
  beams at different wavelengths with low ellipticity}}.
\newblock {\emph{\JournalTitle{J. Opt.}}} \textbf{\bibinfo{volume}{23}},
  \bibinfo{pages}{094001} (\bibinfo{year}{2021}).

\bibitem{basudevpnas}
\bibinfo{author}{Ramaiya, A.}, \bibinfo{author}{Roy, B.},
  \bibinfo{author}{Bugiel, M.} \& \bibinfo{author}{Sch{\"a}ffer, E.}
\newblock \bibinfo{journal}{\bibinfo{title}{Kinesin rotates unidirectionally
  and generates torque while walking on microtubules}}.
\newblock {\emph{\JournalTitle{Proc. Natl. Acad. Sci. U.S.A}}}
  \textbf{\bibinfo{volume}{114}}, \bibinfo{pages}{10894--10899}
  (\bibinfo{year}{2017}).

\bibitem{lokeshjopc}
\bibinfo{author}{Lokesh, M.}, \bibinfo{author}{Vaippully, R.},
  \bibinfo{author}{Bhallamudi, V.~P.}, \bibinfo{author}{Prabhakar, A.} \&
  \bibinfo{author}{Roy, B.}
\newblock \bibinfo{journal}{\bibinfo{title}{Realization of pitch-rotational
  torque wrench in two-beam optical tweezers}}.
\newblock {\emph{\JournalTitle{J. Phys. Commun.}}}
  \textbf{\bibinfo{volume}{5}}, \bibinfo{pages}{115016} (\bibinfo{year}{2021}).

\bibitem{lokeshrscadv}
\bibinfo{author}{Lokesh, M.} \emph{et~al.}
\newblock \bibinfo{journal}{\bibinfo{title}{Estimation of rolling work of
  adhesion at the nanoscale with soft probing using optical tweezers}}.
\newblock {\emph{\JournalTitle{RSC Adv.}}} \textbf{\bibinfo{volume}{11}},
  \bibinfo{pages}{34636--34642} (\bibinfo{year}{2021}).

\bibitem{rahulsoftmatter}
\bibinfo{author}{Vaippully, R.}, \bibinfo{author}{Ramanujan, V.},
  \bibinfo{author}{Gopalakrishnan, M.}, \bibinfo{author}{Bajpai, S.} \&
  \bibinfo{author}{Roy, B.}
\newblock \bibinfo{journal}{\bibinfo{title}{Detection of sub-degree angular
  fluctuations of the local cell membrane slope using optical tweezers}}.
\newblock {\emph{\JournalTitle{Soft Matter}}} \textbf{\bibinfo{volume}{16}},
  \bibinfo{pages}{7606--7612} (\bibinfo{year}{2020}).

\bibitem{pickel2018apparent}
\bibinfo{author}{Pickel, A.~D.} \emph{et~al.}
\newblock \bibinfo{journal}{\bibinfo{title}{Apparent self-heating of individual
  upconverting nanoparticle thermometers}}.
\newblock {\emph{\JournalTitle{Nat. Commun.}}} \textbf{\bibinfo{volume}{9}},
  \bibinfo{pages}{1--12} (\bibinfo{year}{2018}).

\bibitem{rodriguez2001fullprof}
\bibinfo{author}{Rodr{\'\i}guez-Carvajal, J.}
\newblock \bibinfo{journal}{\bibinfo{title}{Fullprof}}.
\newblock {\emph{\JournalTitle{CEA/Saclay, France}}}  (\bibinfo{year}{2001}).

\bibitem{young1993rietveld}
\bibinfo{author}{Young, R.}
\newblock \emph{\bibinfo{title}{The Rietveld method}}, vol.~\bibinfo{volume}{5}
  (\bibinfo{year}{1993}).

\bibitem{magneticbeads}
\bibinfo{author}{Sinha, B.}, \bibinfo{author}{Anandakumar, S.},
  \bibinfo{author}{Oh, S.} \& \bibinfo{author}{Kim, C.}
\newblock \bibinfo{journal}{\bibinfo{title}{Micro-magnetometry for
  susceptibility measurement of superparamagnetic single bead}}.
\newblock {\emph{\JournalTitle{Sens. Actuator A Phys.}}}
  \textbf{\bibinfo{volume}{182}}, \bibinfo{pages}{34--40}
  (\bibinfo{year}{2012}).

\bibitem{hematite1}
\bibinfo{author}{B{\o}dker, F.}, \bibinfo{author}{Hansen, M.~F.},
  \bibinfo{author}{Koch, C.~B.}, \bibinfo{author}{Lefmann, K.} \&
  \bibinfo{author}{M{\o}rup, S.}
\newblock \bibinfo{journal}{\bibinfo{title}{Magnetic properties of hematite
  nanoparticles}}.
\newblock {\emph{\JournalTitle{Phys. Rev. B.}}} \textbf{\bibinfo{volume}{61}},
  \bibinfo{pages}{6826} (\bibinfo{year}{2000}).

\bibitem{hematite2}
\bibinfo{author}{Tadic, M.}, \bibinfo{author}{Panjan, M.},
  \bibinfo{author}{Damnjanovic, V.} \& \bibinfo{author}{Milosevic, I.}
\newblock \bibinfo{journal}{\bibinfo{title}{Magnetic properties of hematite
  ($\alpha$-fe2o3) nanoparticles prepared by hydrothermal synthesis method}}.
\newblock {\emph{\JournalTitle{Appl. Surf. Sci.}}}
  \textbf{\bibinfo{volume}{320}}, \bibinfo{pages}{183--187}
  (\bibinfo{year}{2014}).

\bibitem{nieminen2007optical}
\bibinfo{author}{Nieminen, T.~A.} \emph{et~al.}
\newblock \bibinfo{journal}{\bibinfo{title}{Optical tweezers computational
  toolbox}}.
\newblock {\emph{\JournalTitle{J. Opt. A- Pure. Appl. Op.}}}
  \textbf{\bibinfo{volume}{9}}, \bibinfo{pages}{S196} (\bibinfo{year}{2007}).

\bibitem{gouesbet1990localized}
\bibinfo{author}{Gouesbet, G.}, \bibinfo{author}{Grehan, G.} \&
  \bibinfo{author}{Maheu, B.}
\newblock \bibinfo{journal}{\bibinfo{title}{Localized interpretation to compute
  all the coefficients gnm in the generalized lorenz--mie theory}}.
\newblock {\emph{\JournalTitle{JOSA A}}} \textbf{\bibinfo{volume}{7}},
  \bibinfo{pages}{998--1007} (\bibinfo{year}{1990}).

\bibitem{nieminen2011t}
\bibinfo{author}{Nieminen, T.~A.}, \bibinfo{author}{Loke, V.~L.},
  \bibinfo{author}{Stilgoe, A.~B.}, \bibinfo{author}{Heckenberg, N.~R.} \&
  \bibinfo{author}{Rubinsztein-Dunlop, H.}
\newblock \bibinfo{journal}{\bibinfo{title}{T-matrix method for modelling
  optical tweezers}}.
\newblock {\emph{\JournalTitle{J. Mod. Opt.}}} \textbf{\bibinfo{volume}{58}},
  \bibinfo{pages}{528--544} (\bibinfo{year}{2011}).

\bibitem{refractiveindex}
\bibinfo{author}{Tan, M.~C.}, \bibinfo{author}{Al-Baroudi, L.} \&
  \bibinfo{author}{Riman, R.~E.}
\newblock \bibinfo{journal}{\bibinfo{title}{Surfactant effects on efficiency
  enhancement of infrared-to-visible upconversion emissions of nayf4:yb-er}}.
\newblock {\emph{\JournalTitle{ACS Appl. Mater. Interfaces}}}
  \textbf{\bibinfo{volume}{3}}, \bibinfo{pages}{3910--3915}
  (\bibinfo{year}{2011}).

\bibitem{amrendraEPJ}
\bibinfo{author}{Kumar, A.}, \bibinfo{author}{Gunaseelan, M.},
  \bibinfo{author}{Vaidya, G.}, \bibinfo{author}{Vaippully, R.} \&
  \bibinfo{author}{Roy, B.}
\newblock \bibinfo{journal}{\bibinfo{title}{Estimation of motional parameters
  using emission from upconverting particles optically trapped at the pump
  wavelength}}.
\newblock {\emph{\JournalTitle{Eur. Phys. J. Spec. Top.}}}
  \bibinfo{pages}{1--8}, \doiprefix\url{10.1140/epjs/s11734-021-00399-0}
  (\bibinfo{year}{2021}).

\bibitem{rodriguez2016optical}
\bibinfo{author}{Rodr{\'\i}guez-Sevilla, P.} \emph{et~al.}
\newblock \bibinfo{journal}{\bibinfo{title}{Optical torques on upconverting
  particles for intracellular microrheometry}}.
\newblock {\emph{\JournalTitle{Nano lett.}}} \textbf{\bibinfo{volume}{16}},
  \bibinfo{pages}{8005--8014} (\bibinfo{year}{2016}).

\bibitem{haro2013optical}
\bibinfo{author}{Haro-Gonzalez, P.} \emph{et~al.}
\newblock \bibinfo{journal}{\bibinfo{title}{Optical trapping of nayf 4: Er 3+,
  yb 3+ upconverting fluorescent nanoparticles}}.
\newblock {\emph{\JournalTitle{Nanoscale}}} \textbf{\bibinfo{volume}{5}},
  \bibinfo{pages}{12192--12199} (\bibinfo{year}{2013}).

\bibitem{rollingwithmagnet}
\bibinfo{author}{Alapan, Y.}, \bibinfo{author}{Bozuyuk, U.},
  \bibinfo{author}{Erkoc, P.}, \bibinfo{author}{Karacakol, A.~C.} \&
  \bibinfo{author}{Sitti, M.}
\newblock \bibinfo{journal}{\bibinfo{title}{Multifunctional surface
  microrollers for targeted cargo delivery in physiological blood flow}}.
\newblock {\emph{\JournalTitle{Sci. Robot.}}} \textbf{\bibinfo{volume}{5}},
  \bibinfo{pages}{eaba5726} (\bibinfo{year}{2020}).

\bibitem{bozuyuk2021shape}
\bibinfo{author}{Bozuyuk, U.}, \bibinfo{author}{Alapan, Y.},
  \bibinfo{author}{Aghakhani, A.}, \bibinfo{author}{Yunusa, M.} \&
  \bibinfo{author}{Sitti, M.}
\newblock \bibinfo{journal}{\bibinfo{title}{Shape anisotropy-governed
  locomotion of surface microrollers on vessel-like microtopographies against
  physiological flows}}.
\newblock {\emph{\JournalTitle{Proc. Natl. Acad. Sci. U.S.A.}}}
  \textbf{\bibinfo{volume}{118}}, \bibinfo{pages}{e2022090118}
  (\bibinfo{year}{2021}).

\bibitem{goldnanoheating}
\bibinfo{author}{Honda, M.}, \bibinfo{author}{Saito, Y.},
  \bibinfo{author}{Smith, N.~I.}, \bibinfo{author}{Fujita, K.} \&
  \bibinfo{author}{Kawata, S.}
\newblock \bibinfo{journal}{\bibinfo{title}{Nanoscale heating of laser
  irradiated single gold nanoparticles in liquid}}.
\newblock {\emph{\JournalTitle{Opt. Express}}} \textbf{\bibinfo{volume}{19}},
  \bibinfo{pages}{12375--12383} (\bibinfo{year}{2011}).

\bibitem{photothermaltherapy}
\bibinfo{author}{Jaque, D.} \emph{et~al.}
\newblock \bibinfo{journal}{\bibinfo{title}{Nanoparticles for photothermal
  therapies}}.
\newblock {\emph{\JournalTitle{Nanoscale}}} \textbf{\bibinfo{volume}{6}},
  \bibinfo{pages}{9494--9530} (\bibinfo{year}{2014}).

\bibitem{Kresse1999}
\bibinfo{author}{Kresse, G.} \& \bibinfo{author}{Joubert, D.}
\newblock \bibinfo{journal}{\bibinfo{title}{From ultrasoft pseudopotentials to
  the projector augmented-wave method}}.
\newblock {\emph{\JournalTitle{Phys. Rev. B}}} \textbf{\bibinfo{volume}{59}},
  \bibinfo{pages}{1758--1775} (\bibinfo{year}{1999}).

\bibitem{Kresse1996}
\bibinfo{author}{Kresse, G.} \& \bibinfo{author}{Furthm\"uller, J.}
\newblock \bibinfo{journal}{\bibinfo{title}{Efficient iterative schemes for ab
  initio total-energy calculations using a plane-wave basis set}}.
\newblock {\emph{\JournalTitle{Phys. Rev. B}}} \textbf{\bibinfo{volume}{54}},
  \bibinfo{pages}{11169--11186} (\bibinfo{year}{1996}).

\end{thebibliography}

\end{document}